\documentclass[11pt]{article}
\usepackage{amssymb}
\usepackage{amsmath}
\catcode`@=11
\renewcommand{\section}{\@startsection{section}{1}{0pt}{\medskipamount}
{\medskipamount}{\large\bf}}
\catcode`@=12

\def\th{\theta}

\def\l{\lambda}

\newcommand{\C}{\mathbb C}
\newcommand{\R}{\mathbb R}

\newcommand{\N}{\mathbb N}
\newcommand{\Hcal}{{\cal H}}
\newcommand{\Pcal}{{\cal P}}

\def\N2{$N{=}2$}
\def\pa{\mbox{$\partial$}}
\def\diff{\mbox{d}}
\def\tr{{\rm tr}}
\def\sfrac#1#2{{\textstyle\frac#1#2}}
\def\rd#1{\buildrel{_{_{\hskip 0.01in}\rightarrow}}\over{#1}}
\def\ld#1{\buildrel{_{_{\hskip 0.01in}\leftarrow}}\over{#1}}

\newcommand{\be }{\begin{equation}}
\newcommand{\ee }{\end{equation}}
\newcommand{\adag}{a^{\dagger}}
\newcommand{\cdag}{c^{\dagger}}
\newcommand{\fh}{\hat{f}}
\newcommand{\gh}{\hat{g}}
\newcommand{\Bmunu}{B_{\mu\nu}}
\newcommand{\alf}{\alpha}
\newcommand{\bet}{{\beta}}
\newcommand{\gam}{{\gamma}}

\newcommand{\eps}{\varepsilon}

\newcommand{\alfp}{\alpha^{\prime}}
\newcommand{\psiquer}{\overline{\psi}^{\mu}}
\newcommand{\psiobmu}{\psi^{\mu}}
\newcommand{\psiobnu}{\psi^{\nu}}
\newcommand{\epsb}{\overline{\eps}}
\newcommand{\lb}{\overline{\lambda}}
\newcommand{\Xobmu}{X^{\mu}}
\newcommand{\Xobnu}{X^{\nu}}
\newcommand{\dunalf}{\pa_{\alf}}
\newcommand{\dunbet}{\pa_{\bet}}
\newcommand{\rhoalf}{\rho^{\alf}}

\newcommand{\epsalfbet}{\eps^{\alf\bet}}

\newcommand{\intsig}{\int_{\Sigma}\!\diff^2 \sigma\;}
\newcommand{\Jmunu}{J_{\nu}^{\mu}}
\newcommand{\Gmunu}{G^{\mu\nu}}
\newcommand{\gmunu}{g_{\mu\nu}}
\newcommand{\thetamunu}{\theta^{\mu\nu}}

\newcommand{\zb}{\bar{z}}
\newcommand{\wb}{\bar{w}}

\begin{document}
\begin{titlepage}
\setcounter{page}{0}
\begin{flushright}
hep-th/0103196\\
ITP--UH--03/01\\
\end{flushright}

\vskip 2.0cm

\begin{center}

{\Large\bf  Noncommutative Solitons in Open N=2 String Theory}

\vspace{14mm}

{\large Olaf Lechtenfeld, \ Alexander D. Popov~$^*$ \ and \ Bernd Spendig}
\\[5mm]
{\em Institut f\"ur Theoretische Physik  \\
Universit\"at Hannover \\
Appelstra\ss{}e 2, 30167 Hannover, Germany }\\
{Email: lechtenf, popov, spendig@itp.uni-hannover.de}

\end{center}

\vspace{2cm}

\begin{abstract}
\noindent
Coincident $D2$-branes in open $N{=}2$ fermionic string theory with a
$B$-field background yield an integrable modified $U(n)$ sigma model
on noncommutative $\R^{2,1}$.
This model provides a showcase for an established method
(the `dressing approach') to generate solutions for integrable field
equations, even in the noncommutative case.
We demonstrate the technique by constructing moving $U(1)$ and $U(2)$
solitons and by computing their energies.
It is outlined how to derive multi-soliton configurations 
with arbitrary relative motion; 
they correspond to $D0$-branes moving inside the $D2$-branes.
\end{abstract}

\vfill

\textwidth 6.5truein
\hrule width 5.cm
\vskip.1in

{\small
\noindent ${}^*$
On leave from Bogoliubov Laboratory of Theoretical Physics, JINR,
Dubna, Russia}

\end{titlepage}

\section{Introduction and results}

\noindent
Solitonic solutions of field equations of motion play an essential role
in our understanding of field and string theories beyond perturbation theory.
This persists for the noncommutative extension of scalar and gauge
field theories which have lately come under intense scrutiny because they
appear naturally in string theory~\cite{douglas,cheung,chu,schomerus,ardalan}. 
Indeed, many quantum field theories on a noncommutative space-time are obtained 
in a certain zero-slope limit from effective field theories of open-string 
modes in a particular NS-NS two-form (`$B$-field') background~\cite{seiberg}.
In modern parlance, 
open-string dynamics takes place inside the world volume of $D$-branes 
where a constant $B$-field gets traded for effective noncommutativity 
and a deformed metric. Hence, string theories provide 
a higher-level (commutative) arena for the study of noncommutative field 
theories.

A case in point is the open \N2 fermionic string~\cite{marcus,bischoff,dubna} 
which at tree-level\footnote{
For a one-loop analysis see~\cite{CLN}.}
proved to be identical to self-dual Yang-Mills theory in $2{+}2$ dimensions, 
possibly in the background of a self-dual metric and a K\"ahler 
$B$-field~\cite{OV}.
Turning on the latter in the world volume of $n$ coincident (Kleinian)
$D3$-branes was recently shown to yield noncommutative self-dual $U(n)$
gauge theory~\cite{LPS}.

Starting from $D3$-branes whose world volumes fill the flat (Kleinian) target
space~$\R^{2,2}$, it is interesting to consider lower-dimensional $D$-branes
and describe the effective field theories appearing on their world volumes
in a zero-slope limit. In this paper we show that
open \N2 strings in a $B$-field background induce on the world volume of
$n$ coincident $D2$-branes a noncommutative generalization of a modified $U(n)$
sigma model (with a Wess-Zumino-type term)~\cite{ward}. 
The latter is equivalent to non-relativistic Chern-Simons theory coupled 
to Higgs fields~\cite{dunne}.
For $n$ coincident $D1$-branes we obtain the standard $U(n)$~sigma model in
$1{+}1$ dimensions and its Euclidean version.

By now a lot of solitonic solutions for noncommutative Yang-Mills and 
Yang-Mills-Higgs theories have been constructed~[15--37], 
many of them with the help of the solution-generating technique proposed 
in~\cite{harvey}. 
In the present paper we demonstrate that one can do better 
for `{\it integrable\/}' noncommutative field theories. 
Whenever equations of motion can be written as compatibility conditions
of some {\it linear\/} equations, then there exists a more general technique 
to construct a new solution from an old one, known as the
`{\it dressing\/}' approach \cite{zakharov,zakh2,forgacs}.

We adapt this dressing method to the noncommutative case and
apply it to the noncommutative modified sigma model
in $2{+}1$ dimensions (mentioned above). 
It is outlined how to construct {\it multi\/}-soliton configurations
with arbitrary {\it relative motion\/} of the individual lumps.
To be explicit we illustrate the general scheme by giving examples of
one-soliton solutions for the gauge groups $U(1)$ and $U(2)$, 
including their energies.
Since the model has neither relativistic nor rotational invariance, 
the energy of a noncommutative soliton depends on the direction of its motion. 
For some directions it even falls {\it below\/} the rest energy. 
If all velocities are put to zero one simply obtains static configurations
which are the instantons of the noncommutative two-dimensional 
sigma model. In general, though, our solitons correspond on the string level
to $D0$-branes moving inside $D2$-branes.

\section{Open N=2 strings in a B-field background}

\noindent
We begin by recapitulating some basic facts about the open \N2 
string and how to properly couple it to a $B$-field background. 
The  critical \N2 string lives in 2+2 (real) dimensions, 
i.e. the string world sheet $\Sigma$ is embedded into 
a four-dimensional target space with (Kleinian) signature (2,2).
Its propagating world-sheet degrees of freedom consist of the
embedding coordinates $X^\mu$ (with center-of-mass $x^\mu$) 
plus a set of NSR Majorana spinors $\psi^\mu$, 
with $\mu=1,2,3,4$.
The \N2 string admits couplings 
to a self-dual metric~$g=g_{\mu\nu}(x) \diff{x}^{\mu}\diff{x}^{\nu}$ and a 
K\"ahler NS-NS two-form~$B=\Bmunu(x) \diff{x}^{\mu}{\wedge}\diff{x}^{\nu}$.
We specialize to flat Kleinian space~$\R^{2,2}$ with a metric
\begin{equation}\label{metric}
(g_{\mu\nu})\ =\ \textrm{diag}\,(\xi_1,\xi_1,-\xi_2,-\xi_2)
\qquad\textrm{where}\quad \xi_1,\xi_2 \in \R_+
\end{equation}
are scaling parameters.
In addition, we shall consider $n$~coincident $Dp$-branes for $p=1,2,3$
and gauge to zero all $B$-field components orthogonal to the $D$-brane
world volumes. Concrete expressions for the remaining $B$-field components
shall be given in the next section.

It has been shown~\cite{lind,LPS} that the standard superfield action
in superconformal gauge has to be improved by two $B$-dependent boundary
terms, in order to achieve consistency of the open-string boundary conditions 
with the two residual rigid supersymmetries~\cite{zumino,alvafreedman}
\footnote{
We use $\rho^0=\left(\begin{matrix} 0 & -1 \\ 1 & 0 \end{matrix} \right),\;
\rho^1=\left(\begin{matrix} 0 & 1 \\ 1 & 0 \end{matrix} \right),\;
\{\rho^{\alf},\rho^{\bet}\}=2\eta^{\alf\bet},\;(\eta^{\alf\bet})=
\left(\begin{matrix} -1 & 0 \\ 0 & 1 \end{matrix} \right), \;
\sigma^0=\tau,\; \sigma^1=\sigma.$ \\ \phantom{XX}
Recall that a Majorana spinor $\varphi$ (in 1+1 dimensions) has two components,
and $\overline{\varphi}=\varphi^{\dagger}\rho^0.$
}
\begin{align}
\delta\Xobmu\ &=\ \epsb_1\psi^{\mu}+\Jmunu\epsb_2\psi^{\nu}\quad,
\nonumber\\[4pt]
\delta\psi^{\mu}\ &=\ -i\rhoalf\dunalf\Xobmu\eps_1
+ i \Jmunu\rhoalf\dunalf\Xobnu\eps_2\quad.
\end{align}
Here $(\Jmunu)$ is a constant complex structure compatible with our metric
and $B$-field, i.e.
\begin{align}
\gmunu J_{\lambda}^{\nu}+J_{\mu}^{\nu}g_{\lambda\nu}\ =\ 0
\qquad\mbox{and}\qquad
\Bmunu J_{\lambda}^{\nu}-J_{\mu}^{\nu}B_{\lambda \nu}\ =\ 0\quad .
\end{align}
Its nonzero components are the following: $J_2^1=-J_1^2=J_4^3=-J_3^4=1.$
The consistent gauge-fixed \N2 string action reads~\cite{LPS}
\begin{equation}\label{action2}
S\ =\ -\frac{1}{4\pi\alf^{\prime}}\!\intsig  \big[
(\eta^{\alf\bet}\gmunu+\epsalfbet 2\pi\alfp\Bmunu)\,\dunalf\Xobmu\dunbet\Xobnu
+ (\gmunu+2\pi\alfp\Bmunu)\,i \psiquer\rhoalf\dunalf\psi^{\nu}\big]\quad .
\end{equation}
It is important to note that the action functional~(\ref{action2}) 
cannot be written in terms of superfields.

\section{Branes and Seiberg-Witten limits}\label{seiberg}

\noindent
{\bf General formulae}. 
We now investigate how open \N2 strings are affected by
a background $B$-field on $D$-brane world volumes.
Without loss of generality, we only turn on $B$-field components
parallel to the $D$-brane world volumes.
Furthermore, we switch on only their `magnetic' components, 
in order to allow for the Seiberg-Witten limit 
to yield noncommutative gauge theory. 
 
The starting point is the form of the open-string 
correlators~\cite{schomerus,seiberg},
\begin{align}\label{propa}
\langle \Xobmu(\tau)\,\Xobnu(\tau^{\prime})\rangle\ &=\ 
-\alfp \Gmunu\,\ln(\tau-\tau^{\prime})^2\ +\
\sfrac{i}{2}\thetamunu\,\eps(\tau-\tau^{\prime})\quad , \\[8pt]
\langle \psiobmu(\tau)\ \psiobnu(\tau^{\prime})\rangle\ &=\ 
\frac{\Gmunu}{\tau-\tau^{\prime}}\quad , \label{propa2}
\end{align}
for $\tau,\tau^{\prime}\in \pa\Sigma.$ Here, 
\begin{equation}
[(g+2\pi\alf^{\prime}B)^{-1}]^{\mu\nu}\ =\
\Gmunu\,+\,\frac{\theta^{\mu\nu}}{2\pi\alfp} 
\end{equation}
yields the effective metric $G=(G_{\mu\nu})$ seen by the open string 
and gives rise to the noncommutativity matrix $\theta=(\theta^{\mu\nu})$ 
appearing in $[\Xobmu(\tau),\Xobnu(\tau)]=i\thetamunu$~\cite{schomerus}.
The most general nontrivial `magnetic' field $B=(\Bmunu)$ has components
\begin{equation}\label{bchoice}
B_{12}=-B_{21}=:B_1, \quad B_{34}=-B_{43}=:B_2 \quad ,
\end{equation}
but depending on the dimension and orientation of the $D$-branes 
one or both parameters may be equal to zero.
The inverse open-string metric and the noncommutativity parameters read
\begin{align}
(G^{\mu\nu})\ &=\ \mbox{diag}\,\Bigl(
\frac{\xi_1}{\xi_1^2+(2\pi\alfp B_1)^2}\ ,\ 
\frac{\xi_1}{\xi_1^2+(2\pi\alfp B_1)^2}\ ,\ 
\frac{-\xi_2}{\xi_2^2+(2\pi\alfp B_2)^2}\ ,\ 
\frac{-\xi_2}{\xi_2^2+(2\pi\alfp B_2)^2}\Bigr) \quad, \nonumber \\[10pt]
\theta^{12}\ &=\ -\theta^{21}\ =\ 
-\frac{(2\pi\alfp)^2B_1}{\xi_1^2+(2\pi\alfp B_1)^2} \quad,\qquad
\theta^{34}\ =\ -\theta^{43}\ =\ 
-\frac{(2\pi\alfp)^2B_2}{\xi_2^2+(2\pi\alfp B_2)^2} \quad.
\end{align}
Notice that for $B_2{=}{-}B_1$ the background will be self-dual, 
and the action~(\ref{action2}) will have $N{=}4$ supersymmetry
\cite{alvafreedman}. 

Finally, for the effective open-string coupling $G_s$,
which is related to the closed-string coupling $g_s$ via
$G_s=g_s [{\rm det} G/{\rm det}(g+2 \pi \alfp B)]^{1/2}$,
we obtain
\begin{equation}
G_s\ =\ g_s\Bigl[\bigl(1+(2\pi\frac{\alf^{\prime}}{\xi_1}B_1)^2\bigr)
\bigl(1+(2\pi\frac{\alf^{\prime}}{\xi_2}B_2)^2\bigr)\Bigr]^{1/2}\ =\
(2\pi)^{2-p}\,(\alf^{\prime})^{\frac{3-p}{2}}\,g_{\textrm{YM}}^2 \quad,
\end{equation}
including the relation to the Yang-Mills coupling~$g_{\textrm{YM}}$ 
on the $Dp$-brane~\cite{seiberg}.

\noindent
{\bf D3-branes}.
Consider $n$ coincident (Kleinian) $D3$-branes with world-volume signature 
$(++-\,-)$. Then the Seiberg-Witten limit consists of taking $\alfp{\to}0$ 
while sending $\xi_1=\xi_2 =(\alfp)^2 \to 0$ (and therefore $\gmunu \to 0$) 
so that $G,\; G^{-1}$, and $\theta$ remain finite. 
Let us now rescale coordinates, 
\begin{equation}
x^1\ \to\ 2\pi B_1 x^1 \quad, \qquad 
x^2\ \to\ 2\pi B_1 x^2 \quad, \qquad 
x^3\ \to\ 2\pi B_2 x^3 \quad, \qquad 
x^4\ \to\ 2\pi B_2 x^4 \quad,
\end{equation} 
and denote them by the same letter $x^{\mu}$. 
Then, the components of $G$ and $\theta$ must be transformed as befits
rank-two tensors. Explicitly, we have
\begin{align} \label{d3}
(G^{\mu\nu})\ &\to\ (\eta^{\mu\nu})\ =\ \mbox{diag}\,(+1,+1,-1,-1) 
\quad, \nonumber \\[10pt]
\theta^{12}\ =\ -\theta^{21}\ &\to\ -4\pi^2 B_1 \quad, \qquad
\theta^{34}\ =\ -\theta^{43}\  \to\ -4\pi^2 B_2 \quad.
\end{align}  
Thus, in these coordinates the commutative limit $\theta \rightarrow 0$ 
corresponds to $B \rightarrow 0.$

\noindent
{\bf D2-branes}.
Next we have a look at $n$ coincident (Minkowskian) $D2$-branes with 
world volumes extending along 
$x^1,x^2,x^4.$ In this case we take $\xi_1=(\alfp)^2$ but $\xi_2=1$ and in
the Seiberg-Witten limit send $\alfp \to 0.$ 
Our subsequent rescaling for this case,
\begin{equation}
x^1\ \to\ 2\pi B_1 x^1 \quad, \qquad
x^2\ \to\ 2\pi B_1 x^2 \quad, \qquad
x^3\ \to\ x^3 \quad, \qquad
x^4\ \to\ x^4 \quad,
\end{equation}
sends the inverse open-string metric and the noncommutativity parameter to
\begin{equation} \label{d2}
(G^{\mu\nu})\ \to\ (\eta^{\mu\nu})\ =\ \mbox{diag}\,(+1,+1,-1,-1)
\quad, \qquad
\theta^{12}\ =\ -\theta^{21}\ \to\ -4\pi^2 B_1\ =:\ \theta
\end{equation}
(remembering that $\theta^{34}{=}0$ in this case).
It is obvious that the $D2$-brane results can be obtained from the $D3$-brane 
equations by simply setting $B_2=0$. Note that by considering 
Euclidean $D1$-branes we arrive at the same formulae (\ref{d2}).

\section{Effective field theories on D-brane world volumes}\label{effective}
\noindent
Eleven years ago it was discovered~\cite{OV} that the open \N2 fermionic
string at tree level is identical to {\it self-dual\/} Yang-Mills field theory
in $2{+}2$ dimensions. The complete absence of a massive physical spectrum
ties in with the vanishing of all string amplitudes beyond three-point.
In \cite{LPS} it was shown that turning on a non-vanishing constant
$B$-field background yields {\it noncommutative\/} self-dual Yang-Mills 
as the effective field theory in the target. 
We shall now extend our analysis to the case where the $B$-field is 
restricted to the world volume of lower-dimensional $D$-branes. 

Apart from replacing $g\to G$ in the propagators 
(\ref{propa}) and~(\ref{propa2}), 
turning on a $B$-field merely multiplies
each primitive open-string amplitude, obtained at $\theta{=}0$
from a disk diagram with a fixed leg ordering $(1,2,\ldots,M)$, by a factor of
\vspace{-0.3cm}
\begin{equation}\label{phase}
\prod_{1\le j<\ell\le M}
\exp \,\bigl\{-{i\over2}k_{j\mu}\,\th^{\mu\nu}\,k_{\ell\nu}\bigr\} \quad,
\end{equation}
where $k_j$ denotes the momentum of the $j$th external leg.
As delineated in~\cite{seiberg} the effect of the phase~(\ref{phase})
is subsumed in the effective world-volume field theory by replacing
the ordinary product of fields with the Moyal (star) product.

The \N2 string is no exception to this rule~\cite{LPS}:
Its tree-level amplitudes on the space-time filling brane are
reproduced by the noncommutative extension of the scalar field theory
for a {\it prepotential\/} of {\it self-dual\/} $U(n)$ gauge theory.
The choice of prepotential is a matter of gauge. In fact, by choosing
a suitable Lorentz frame we can always arrive at Leznov's gauge~\cite{L}
or at Yang's gauge~\cite{Y}
and even scale the Yang-Mills coupling~$g_{\textrm{YM}}$ to unity~\cite{LS}.
Since the Leznov as well as the Yang gauge break $SO(2,2)$ invariance, the 
resulting scalar field theories and equations have a reduced global symmetry.
In the following we shall employ both gauges and restrict to 
lower-dimensional $D$-branes by setting appropriate $B$-field components
and coordinate dependencies to zero.

\noindent
{\bf D3-branes.}
This is the maximal case already discussed in~\cite{LPS}.
The cubic Lagrangian for the $u(n)$-valued Leznov prepotential~$\phi$ reads
\begin{equation} \label{d3lag}
L\ =\ \sfrac{1}{2}\eta^{\mu\nu}\tr\; \pa_{\mu}\phi\star\pa_{\nu}\phi +
\sfrac{1}{3}\tr\;\phi\star\bigl[(\pa_2+\pa_4)\phi\star(\pa_1-\pa_3)\phi -
(\pa_1-\pa_3)\phi\star(\pa_2+\pa_4)\phi\bigr] \quad,
\end{equation}
where `$\tr$' traces over the $u(n)$ algebra and fields are multiplied
by means of the noncommutative star product
\vspace{-0.3cm}
\begin{equation}
(f \star g)(x)\ =\ f(x)\,\exp\,\bigl\{ {i\over2} 
{\ld{\partial}}_\mu \,\theta^{\mu\nu}\, {\rd{\partial}}_\nu \bigr\}\,g(x)\quad.
\end{equation}

\noindent
{\bf D2-branes.}
For the case of coincident $D2$-branes our reasoning leads
to the Lagrangian
\begin{equation} \label{d2lag}
L\ =\ \sfrac{1}{2}\eta^{ab}\tr\;\pa_a \phi\star\pa_b \phi +
\sfrac{1}{3}\tr\;\phi\star\bigl[(\pa_2+\pa_4)\phi\star\pa_1\phi -
\pa_1\phi\star(\pa_2+\pa_4)\phi\bigr] \quad,
\end{equation} 
where $a,b,\ldots=1,2,4$. 
We see that in accordance with the
discussion above the Lagrangian (\ref{d2lag}) can be obtained from 
(\ref{d3lag}) upon reducing from the $D3$-brane world volume to a $D2$-brane 
world volume by imposing $\pa_3\phi=0$.
Our choice of $B$-field then implies (see (\ref{d2})) that the time 
coordinate~$x^4$ becomes commutative.
{}From now on, we shall relabel our coordinates as follows,
\begin{equation}
x\ =\ x^1 \quad,\qquad
y\ =\ x^2 \quad,\qquad
t\ =\ -x^4\quad.
\end{equation}

The Lagrangian (\ref{d2lag}) produces a Leznov-type equation of motion,
\begin{equation}\label{motion2}
\pa_x^2\phi-\pa_u\pa_v\phi+\pa_v\phi\star\pa_x\phi-\pa_x\phi\star\pa_v\phi
\ =\ 0 \quad,
\end{equation}
where we defined
\begin{equation}
u\ :=\ \sfrac{1}{2}(t+y)\quad,\qquad 
v\ :=\ \sfrac{1}{2}(t-y)\quad,\qquad 
\pa_u\ =\ \pa_t+\pa_y\quad,\qquad 
\pa_v\ =\ \pa_t-\pa_y\quad.
\end{equation}

\noindent
{\bf D1-branes.}
By considering open \N2 strings interacting with $n$ coincident $D1$-branes 
extending in the $x$-direction we obtain in the Seiberg-Witten limit 
(see section~\ref{seiberg}) an effective field theory on the $D1$-brane 
world volume~$\R^{1,1}$, with the following equation of motion:
\begin{equation}\label{d1a}
\pa_x^2\phi-\pa_t^2\phi+\pa_t\phi\star\pa_x\phi-\pa_x\phi\star\pa_t\phi
\ =\ 0 \quad.
\end{equation}
Alternatively, 
for Euclidean $D1$-branes the effective scalar field on $\R^{2,0}$ obeys
\begin{equation}\label{d1b}
\pa_x^2\phi+\pa_y^2\phi+\pa_x\phi\star\pa_y\phi-\pa_y\phi\star\pa_x\phi
\ =\ 0 \quad.
\end{equation} 
It is easy to see that equations (\ref{d1a}) and (\ref{d1b}) can be 
obtained from equation (\ref{motion2}) by imposing the conditions 
$\pa_y \phi=0$ and $\pa_t\phi=0$, respectively.

\noindent
{\bf Operator formalism.}
In the remainder of this paper we concentrate on the noncommutative field 
theory (\ref{d2lag}) in $2{+}1$ dimensions, which can be obtained in a 
Seiberg-Witten limit of the open \N2 string with $n$ coincident
$D2$-branes and a constant $B$-field background. 
It is well known that the star product, e.g. in~(\ref{d2lag}),
can be dropped by passing to noncommutative coordinates, 
$x^\mu\to\hat{x}^\mu$, satisfying $[\hat{x}^\mu,\hat{x}^\nu]=i\theta^{\mu\nu}$.
In our $D2$-brane context we have $(t,x,y)\to(t,\hat{x},\hat{y})$ where
\begin{equation}
[t,\hat{x}]\ =\ [t,\hat{y}]\ =\ 0\quad, \qquad 
[\hat{x},\hat{y}]\ =\ i\theta \qquad\textrm{with}\quad 
\theta = -4\pi^2 B_1 > 0 \quad .
\end{equation}
The metric $(\eta_{ab})=\textrm{diag}(+1,+1,-1)$ and the parameter $\theta$ 
are extracted from equation (\ref{d2}).

Using complex noncommutative coordinates 
$\hat{z}=\hat{x}+i\hat{y}$ and $\hat{\zb}= \hat{x}-i\hat{y}$ 
we introduce creation and annihilation operators 
\begin{equation}
a\ =\ \frac{1}{\sqrt{2\theta}}\,\hat{z} \qquad\textrm{and}\qquad 
\adag\ =\ \frac{1}{\sqrt{2\theta}}\,\hat{\zb} \qquad \textrm{so that}\quad 
[a,\adag]\ =\ 1 \quad. 
\end{equation}
They act on the Fock space $\Hcal$ with an orthonormal basis 
$\{|n\rangle,\,n=0,1,2,\ldots\}$ such that 
\begin{equation}
a\,|0\rangle\ =\ 0\quad, \qquad 
a\,|n\rangle\ =\ \sqrt{n}\,|n{-}1\rangle \quad, \qquad 
\adag|n\rangle\ =\ \sqrt{n{+}1}\,|n{+}1\rangle \quad .
\end{equation}  
Then, any function $f(t,z,\zb)$ can be related to an operator-valued 
function $\fh(t)$ acting in $\Hcal$, 
with the help of the Moyal-Weyl map (see e.g. \cite{alv,gross3})
\begin{equation}
f(t,z,\zb)\quad \longrightarrow \quad \fh(t)\ =\ 
\int\!\frac{2i\,\diff{p}\,\diff{\bar{p}}}{(2\pi)^2}\; \tilde{f}(t,p,\bar{p})\;
e^{-i\sqrt{2\theta}(\bar{p}a+p \adag)} \quad,
\end{equation}
where $\tilde{f}(t,p,\bar{p})$ is the Fourier transform of $f(t,z,\zb)$ 
with respect to $(z,\zb)$. Under this map, we have
$f{\star}g \to \fh\,\gh$ and
\begin{equation}
\int\! \diff{x}\,\diff{y}\,f\quad \longrightarrow \quad
2\pi \theta \,\mbox{Tr}\, \fh\ =\ 
2\pi \theta \sum_{n \geq 0} \langle n|\fh |n \rangle \quad,
\end{equation}
where `Tr' signifies the trace over the Fock space~$\Hcal$.
The operator formulation turns spatial derivatives into commutators,
\begin{equation} \label{opbrac}
\pa_x f \quad \longrightarrow \quad \frac{i}{\theta}\, [\hat{y},\fh] 
\ =:\ \hat{\pa}_x \fh
\qquad\textrm{and}\qquad 
\pa_y f \quad \longrightarrow \quad -\frac{i}{\theta}\, [\hat{x},\fh] 
\ =:\ \hat{\pa}_y \fh \quad.
\end{equation}
For notational simplicity we will from now on omit the hats over the operators.

In the operator language, the Leznov-type equation (\ref{motion2}) 
may be rewritten as follows,
\begin{equation}\label{motion2a}
\pa_x^2\phi -\pa_u\pa_v\phi +[\pa_v \phi\,,\, \pa_x \phi]\ =\ 0 \quad ,
\end{equation}
where the elements of the $u(n)$ matrices $\phi$ are $t$-dependent operators 
in the Fock space $\Hcal$, and derivatives 
$\pa_x$, $\pa_u=\pa_t+\pa_y$ and $\pa_v=\pa_t-\pa_y$
are understood as in (\ref{opbrac}).

\section{Noncommutative solitons in 2+1 dimensions}\label{noncomm}
\noindent
In the previous sections we have laid the groundwork for the main
part of this paper, which sets out to construct solitonic classical
configurations for the integrable effective field theory of open \N2 strings
on coincident noncommutative $D2$-branes. 
The crucial ingredient for solving the noncommutative field equations
is the `dressing method', which was invented to generate solutions for 
commutative integrable systems~\cite{zakharov,zakh2,forgacs}
and is easily extended to the noncommutative setup. 
The purpose of this section is, firstly, to present this important technique
in its generality and adapt it to the noncommutative situation before,
secondly, demonstrating its power by deriving one-soliton configurations
for the $U(1)$ and $U(2)$ gauge groups and calculating their energies.

\noindent
{\bf Linear system.}
Let us consider the two linear equations 
\begin{equation}\label{linsys}
(\zeta \pa_x -\pa_u)\psi\ =\ A\,\psi \qquad\textrm{and}\qquad
(\zeta \pa_v -\pa_x)\psi\ =\ B\,\psi \quad,
\end{equation}
which can be obtained from the Lax pair for the self-dual Yang-Mills equations 
in $\R^{2,2}$ \cite{ivle} by gauge-fixing and imposing the condition 
$\pa_3 \psi =0$.
Here, $\psi$ depends on $(t,x,y,\zeta)$ or, equivalently, on $(x,u,v,\zeta)$ 
and is an $n{\times}n$ matrix whose elements act as operators in the Fock 
space $\Hcal$. The matrices $A$ and~$B$ are of the same type as $\psi$ but
do not depend on~$\zeta$. 
The spectral parameter~$\zeta$ lies in the extended complex plane.
The matrix $\psi$ is subject to the following 
reality condition~\cite{ward}:
\begin{equation}\label{real}
\psi(t,x,y,\zeta)\;[\psi(t,x,y,\bar{\zeta})]^{\dagger}\ =\ 1 \quad,
\end{equation}
where `$\dagger$' is hermitian conjugation. We also impose on $\psi$ the 
standard asymptotic conditions \cite{ivle}
\begin{align}
\psi(t,x,y,\zeta\to\infty)\ &=\ 1\ +\ \zeta^{-1}\phi(t,x,y)\ +\ O(\zeta^{-2})
\quad, \label{asymp1} \\[4pt]
\psi(t,x,y,\zeta\to0)\ &=\ \Phi^{-1}(t,x,y)\ +\ O(\zeta)\quad. \label{asymp2}
\end{align}
As we shall see in a moment,
it is no coincidence that the letter~$\phi$ in (\ref{asymp1}) is also used
to denote the (operator-valued) Leznov field in~(\ref{motion2a}).

\noindent
{\bf Compatibility conditions.}
The compatibility conditions for the linear system of differential equations
(\ref{linsys}) are
%\vspace{-0.5cm}
\begin{align}
\pa_x B -\pa_v A\ =\ 0 \quad ,\label{comp1} \\[4pt]
\pa_x A -\pa_u B -[A,B]\ =\ 0 \quad . \label{comp2}
\end{align}
The solution of (\ref {comp1}) is 
%\vspace{-0.5cm}
\begin{equation} \label{sol1}
A\ =\ \pa_x\phi \qquad \textrm{and} \qquad B\ =\ \pa_v \phi \quad,
\end{equation}
where $\phi$ is the same matrix as in (\ref{asymp1}). By substituting 
(\ref{sol1}) into (\ref{comp2}) we reproduce the second-order equation 
(\ref{motion2a}), which identifies $\phi$ as the Leznov field. 
 
Alternatively, one can first solve (\ref{comp2}) via
\begin{equation}\label{sol2}
A\ =\ \Phi^{-1}\pa_u\Phi\qquad\textrm{and}\qquad B\ =\ \Phi^{-1}\pa_x\Phi\quad,
\end{equation}
where $\Phi^{-1}$ is the matrix in (\ref{asymp2}). By inserting
(\ref{sol2}) into the remaining condition (\ref{comp1}) we gain an equivalent 
(Yang-type) form of our field equation~(\ref{motion2a}), namely
\begin{equation} \label{yangtype}
\pa_x(\Phi^{-1}\pa_x \Phi )-\pa_v(\Phi^{-1}\pa_u \Phi)\ =\ 0 \quad .
\end{equation}
Returning to the original coordinates,
\begin{equation} \label{yangtype2}
\pa_x(\Phi^{-1}\pa_x \Phi ) + \pa_y(\Phi^{-1}\pa_y \Phi ) -
\pa_t(\Phi^{-1}\pa_t \Phi ) + 
\pa_y(\Phi^{-1}\pa_t \Phi ) - \pa_t(\Phi^{-1}\pa_y \Phi ) \ =\ 0 \quad,
\end{equation}
one notices that this equation is not $SO(2,1)$ Lorentz invariant.
However, it is the integrable equation of motion for 
a modified sigma model in $2{+}1$ dimensions~\cite{ward}.
Finally we remark that (\ref{sol1}) and (\ref{sol2}) also follow directly from 
(\ref{linsys}) and the asymptotic forms (\ref{asymp1}) and (\ref{asymp2}).

\noindent
{\bf Solutions.}
In order to generate solutions of the equations (\ref{motion2a}) and 
(\ref{yangtype}) we employ the so-called `dressing method'
\cite{zakharov,zakh2,forgacs,ward}. 
More concretely, we look for solutions to the system~(\ref{linsys}) 
with (\ref{real}) of the form 
\begin{equation}\label{movsol}
\psi(t,x,y,\zeta)\ =\ 
1\ +\ \sum_{k=1}^{m}\frac{\mu_k-\bar{\mu}_k}{\zeta-\mu_k}\, P_k(t,x,y) \quad,
\end{equation}
where the $\mu_k$ are complex constants 
(usually Im $\mu_k<0$ for all~$k$)
and the matrices $P_k$ are independent
of $\zeta.$ In the commutative case such $\psi$ describe $m$ 
{\it moving\/} solitons which scatter trivially and whose 
velocities are determined by the constants $\mu_k.$
Subjecting the ansatz (\ref{movsol}) to the reality condition (\ref{real}) 
we find that the $P_k$ have to satisfy certain algebraic equations.
An obvious solution to the latter is found by taking the $P_k$ to
be mutually orthogonal hermitian projectors:
\begin{equation}\label{proj}
P_k^{\dagger}\ =\ P_k \quad, \qquad P_k^2\ =\ P_k \quad, \qquad 
P_{k_1}P_{k_2}\ =\ 0 \quad \textrm{for} \quad k_1\neq k_2 \quad . 
\end{equation}

{}From (\ref{linsys}) and (\ref{real}) we learn that 
\begin{align}
A(t,x,y)\ =\ 
(\zeta\pa_x\psi-\pa_u\psi)(t,x,y,\zeta)\;[\psi(t,x,y,\bar{\zeta})]^{\dagger}
\quad , \label{A1} \\[4pt]
B(t,x,y)\ =\ 
(\zeta\pa_v\psi-\pa_x\psi)(t,x,y,\zeta)\;[\psi(t,x,y,\bar{\zeta})]^{\dagger}
\quad \label{B1}.
\end{align}
The matrices $A$ and $B$ do not depend on $\zeta$; therefore the poles at
$\zeta=\mu_k$ on the r.h.s. of (\ref{A1}) and (\ref{B1}) have to be removable. 
This leads to the following differential equations for the projectors $P_k$:
\begin{equation}\label{Pdiff}
(\mu_k \pa_x P_k-\pa_u P_k) \Bigl( 1-\sum_{\ell=1}^m
\frac{\mu_\ell{-}\bar{\mu}_\ell}{\mu_k{-}\bar{\mu}_\ell} P_\ell \Bigr)
\ =\ 0 \ =\
(\mu_k \pa_v P_k-\pa_x P_k) \Bigl( 1-\sum_{\ell=1}^m
\frac{\mu_\ell{-}\bar{\mu}_\ell}{\mu_k{-}\bar{\mu}_\ell} P_\ell \Bigr) \quad.
\end{equation} 
If $r_k$ is the matrix rank of $P_k$  and $n{\ge}2$ 
we can always take $P_k$ to be of the form
\begin{equation}\label{Pform}
P_k\ =\ T_k\,(T_k^{\dagger} T_k)^{-1}\,T_k^{\dagger} 
\qquad\textrm{for}\quad k=1,\ldots,m \quad ,
\end{equation}
where the $T_k$ are arbitrary $n\times r_k$ matrices 
subject to $T_k^{\dagger} T_\ell=0$ for $k\neq\ell$. 
As before, the elements of $T_k, T_k^{\dagger}$ and $P_k$ are Fock-space
operators.
Substituting (\ref{Pform}) into (\ref{Pdiff}) 
we find that the equations (\ref{Pdiff}) hold if
the $T_k^{\dagger}$ satisfy
\begin{align}\label{Tdiff}
\mu_k \pa_x T_k^{\dagger}-\pa_u T_k^{\dagger}\ =\ 0 \qquad\textrm{and}\qquad 
\mu_k \pa_v T_k^{\dagger}-\pa_x T_k^{\dagger}\ =\ 0 \quad .
\end{align}  
The {\it general\/} solution of these equations is 
$T_k^{\dagger}=T_k^{\dagger}(\bar{w}_k)$, 
meaning the $T_k$ are arbitrary functions of 
\begin{equation}\label{gensol}
w_k\ :=\ x+\bar{\mu}_k u +\bar{\mu}_k^{-1} v\ =\ x + 
\sfrac{1}{2}(\bar{\mu}_k-\bar{\mu}_k^{-1})\,y +
\sfrac{1}{2}(\bar{\mu}_k+\bar{\mu}_k^{-1})\,t \quad.
\end{equation}
Hence, if the matrix elements of $T_k$ are arbitrary holomorphic functions 
of the complex linear combination $w_k$ of the coordinates $t,x,y$, then 
the projectors (\ref{Pform}) satisfy the differential equations~(\ref{Pdiff}). 

The sub-ansatz (\ref{proj}) is rather restrictive for describing
{\it multi\/}-soliton configurations. Those are captured better by putting
(cf.~\cite{forgacs,ward})
\begin{equation}\label{Pform2}
P_k\ =\ \sum_{\ell=1}^m \, T_\ell\,
\frac{\Gamma^{\ell k}}{\mu_k{-}\bar{\mu}_k}\,T_k^\dagger
\qquad\textrm{for}\quad k=1,\ldots,m \quad ,
\end{equation}
where the $T_k(t,x,y)$ are $n\times r$ matrices, now unconstrained, 
and $\Gamma^{\ell k}(t,x,y)$ are $r\times r$ matrices 
with $r\ge1$, so that
\begin{equation}\label{movesol2}
\psi(t,x,y,\zeta)\ =\
1\ +\ \sum_{k,\ell=1}^{m}\frac{1}{\zeta-\mu_k}\, 
T_\ell\; \Gamma^{\ell k}\, T_k^\dagger \quad.
\end{equation}
Substituting (\ref{movesol2}) into (\ref{real}) we learn that
the $\Gamma^{\ell k}$ must invert the $r\times r$ matrices
\begin{equation}\label{Gamma}
\widetilde{\Gamma}_{kp}\ =\ \frac{1}{\mu_k{-}\bar{\mu}_p}\,T_k^\dagger\,T_p
\quad, \qquad\textrm{i.e.}\qquad
\sum_{k=1}^m\,\Gamma^{\ell k}\,\widetilde{\Gamma}_{kp}\ =\ 
\delta_{\ p}^\ell \quad.
\end{equation}
Demanding again vanishing residues for the poles at $\zeta=\mu_k$ 
on the r.h.s. of (\ref{A1}) and (\ref{B1}) 
we obtain the same differential equations as before, namely~(\ref{Tdiff}), 
but now for the $T_k^\dagger$ in the ansatz~(\ref{movesol2}). 
Thus, again every collection of holomorphic matrix functions $T_k(w_k)$ 
provides a solution.

We observe that plugging the ansatz (\ref{movsol}) into the
formulae (\ref{asymp1}) and (\ref{asymp2}) reveals the solutions
\begin{equation}\label{explsol}
\phi\ =\ \sum_{k=1}^m(\mu_k-\bar{\mu}_k)\,P_k
\qquad\textrm{and}\qquad
\Phi^{-1}\ =\ \Phi^{\dagger}\ =\
1\ -\ \sum_{k=1}^m\frac{\mu_k-\bar{\mu}_k}{\mu_k}\,P_k
\end{equation}
of equations (\ref{motion2a}) and (\ref{yangtype}) in terms of projectors.
In the case of (\ref{Pform2}) the solutions take the form
\begin{equation}\label{explsol2}
\phi\ =\ \sum_{k,\ell=1}^m T_\ell\; \Gamma^{\ell k}\, T_k^\dagger
\qquad\textrm{and}\qquad
\Phi^{-1}\ =\ \Phi^{\dagger}\ =\
1\ -\ \sum_{k,\ell=1}^m \frac{1}{\mu_k}\,T_\ell\;\Gamma^{\ell k}\,T_k^\dagger
\quad.
\end{equation}
Note that formally the noncommutativity never entered our considerations.

\noindent
{\bf Solitons.} 
Formulae (\ref{explsol}) and (\ref{explsol2}) represent two rather general 
classes of explicit solutions to our Leznov- and Yang-type field equations
(\ref{motion2a}) and  (\ref{yangtype}).
As solutions of the reduced noncommutative self-dual Yang-Mills equations
they are of BPS character. 
If we want to specialize to {\it solitons\/} (i.e. moving lumps of energy), 
we should impose the condition that the energy of such field configurations 
is finite. For this consideration we find it convenient to switch from
the Leznov- to the Yang-type description, which makes use of the
$U(n)$-valued field~$\Phi(t,x,y)$.
The computation of the energy gets quite involved already in the 
commutative limit when the soliton configuration~$\Phi$ is 
time-dependent~\cite{ward,sutcliffe,ioan1,ioan2},
and even more so for {\it multi\/}-soliton configurations.
To keep the algebra manageable we concentrate on {\it one\/}-soliton solutions
for the gauge groups $U(1)$ and $U(2)$.

Putting $m{=}1$ and dropping the index we get
$\Gamma=\frac{\mu{-}\bar{\mu}}{T^\dagger T}$, and 
the solutions (\ref{explsol}) and (\ref{explsol2}) simplify to
\begin{equation}\label{onedimsol}
\phi\ =\ (\mu-{\bar \mu})\,P \qquad\textrm{and}\qquad
\Phi\ =\ 1 +\,\frac{\mu-{\bar\mu}}{{\bar \mu}}\,P \quad.
\end{equation}
It is assumed that $\mu$ is not real.
Furthermore, we have $\pa_x=\pa_w+\pa_{\wb}$ and
\begin{align}
\pa_t\ =\ \sfrac12(\bar{\mu}+\bar{\mu}^{-1})\pa_w +\sfrac12(\mu+\mu^{-1}) 
\pa_{\wb} \quad, \nonumber \\[4pt]
\pa_y\ =\ \sfrac12(\bar{\mu}-\bar{\mu}^{-1})\pa_w +\sfrac12(\mu-\mu^{-1})
\pa_{\wb} \quad,
\end{align}
so that (\ref{Pdiff}) reduces to 
\begin{equation}\label{reduced}
(\pa_w P)\,(1-P)\ =\ 0\qquad\Longrightarrow\qquad(1-P)\;\pa_{\wb}P\ =\ 0\quad.
\end{equation}

Recall that our solutions $\Phi$ depend on $t,x,y$ only through $w$ and $\wb$.
It is therefore convenient to switch from $z,\zb$ 
(as used in section~\ref{effective})
to the `co-moving' coordinates $w,\wb$.
For the latter we have (in the operator formalism)
\begin{equation}
[w,\wb]\ =\ \sfrac{i}{2}\theta\,(\mu-\bar{\mu}-\mu^{-1}+\bar{\mu}^{-1})
\ =:\ 2\beta \ >\ 0 \qquad\textrm{for}\quad \textrm{Im}\,\mu <0 \quad.
\end{equation}
Notice that $w\equiv z$ and $\bet=\theta$ when $\mu=-i$ ({\it static\/} case). 
We introduce creation and annihilation operators $\cdag$ and $c$ 
together with a Fock space $\Hcal_t$ related to the `co-moving' 
coordinates $w,\wb$ via
\begin{align}\label{ccreat}
c\ =\ \frac{1}{\sqrt{2\bet}}\,w \quad, \qquad 
\cdag\ =\ \frac{1}{\sqrt{2\bet}}\,\wb \quad, \qquad 
|n\rangle_t\ :=\ \frac{(\cdag)^n}{\sqrt{n!}}\,|0\rangle_t\quad,
\phantom{XXXXt}\nonumber\\[4pt]
c\,|0\rangle_t\ =\ 0 \quad, \qquad 
c\,|n\rangle_t\ =\ \sqrt{n}\,|n{-}1\rangle_t \quad, \qquad 
\cdag\,|n\rangle_t\ =\ \sqrt{n{+}1}\,|n{+}1\rangle_t \quad .
\end{align}
Coordinate derivatives are then represented in the standard fashion as
\cite{alv,gross3}
\begin{equation}\label{ccreat2}
\sqrt{2\bet}\,\pa_w \quad\longrightarrow\quad -[\cdag, .\,] \quad, \qquad
\sqrt{2\bet}\,\pa_{\wb} \quad\longrightarrow\quad [c, .\,] \quad. 
\end{equation} 

In the case of commutative space-time coordinates 
there are no smooth soliton solutions
with fields taking values in the abelian $U(1)$ group or its Lie algebra
since then equations (\ref{motion2a}) and (\ref{yangtype}) will be linear. 
For a noncommutative space, though, this is not the case, as has been widely 
observed in the literature. 
In order to solve (\ref{reduced}) for the $U(1)$ case,
one may take as $P$ simply the projector 
\begin{equation}
P_0\ =\ |0\rangle_t \langle 0|_t
\end{equation}
onto a one-dimensional subspace of the Fock space $\Hcal_t.$ 
It is not difficult to see that $P_0$ satisfies the condition~(\ref{reduced}). 
Thus, configurations~(\ref{onedimsol}) with $P{=}P_0$ 
fulfil equations (\ref{motion2a}) and (\ref{yangtype}), respectively. 

Let us advance to a one-soliton solution for the group $U(2)$. 
In this situation, we have $n=2$, and our $r{=}1$ projector~$P$ is constructed 
from a $2{\times}1$ matrix~$T$. As a simple case we take
\begin{equation}\label{toned}
T\ =\ \left(\begin{matrix} \lambda \\ c \end{matrix}\right)
 \qquad\textrm{with}\quad \l\in\C\setminus\{0\}
\end{equation} 
and create the projector 
\begin{align}\label{mat}
P\ =\ T\frac{1}{T^{\dagger}T}T^{\dagger}\ =\ \left(\begin{matrix}
\frac{\l\lb}{N+\l\lb} & \frac{\l}{N+\l\lb}\, \cdag \\[8pt]
c\, \frac{\lb}{N+\l\lb} & c\, \frac{1}{N+\l\lb}\, \cdag \end{matrix}\right)
\qquad\textrm{where}\quad
N\ :=\ \cdag c \quad,
\end{align}
which satisfies (\ref{reduced}). Formulae~(\ref{onedimsol}) now provide
solutions of the field equations (\ref{motion2a}) and (\ref{yangtype}).

\noindent
{\bf Energies.}
As in the commutative case \cite{ward}, the model (\ref{d2lag}) has a
conserved energy-momentum density,
%\vspace{-0.4cm}
\begin{equation}
\Pcal_a\ =\ (\delta^b_a \delta_0^c-\sfrac{1}{2}\eta_{a0}\eta^{bc})
\,\tr\,\pa_b \Phi^{\dagger} \star \pa_c \Phi \quad,
\end{equation}
which yields the energy functional
\begin{align}\label{energy}
E\ &=\ \int\! \diff{x}\,\diff{y} \;\Pcal_0\ =\
\frac{1}{2}\int\!\diff{x}\,\diff{y}\;\tr\,
\Bigl(\pa_t \Phi^{\dagger} \star \pa_t \Phi \,+\,
 \pa_x \Phi^{\dagger} \star \pa_x \Phi \,+\,
 \pa_y \Phi^{\dagger} \star \pa_y \Phi \Bigr) \nonumber\\[8pt]
&=\ \frac{1}{4}\int\!\diff{x}\,\diff{y}\;\tr \biggl[
(\bar{\mu}+\bar{\mu}^{-1})^2\,\pa_w \Phi^{\dagger}\star\pa_w \Phi\ +\
(\mu+\mu^{-1})^2\,\pa_{\wb}\Phi^{\dagger}\star\pa_{\wb}\Phi
\nonumber \\ 
&\phantom{XXXXXXXt} +\,(2+\mu\bar{\mu}+\mu^{-1}\bar{\mu}^{-1})\,
(\pa_w\Phi^{\dagger}\star\pa_{\wb}\Phi\,+\,
\pa_{\wb}\Phi^{\dagger}\star\pa_w\Phi)
\biggr] \quad,
\end{align}
where the last expression is valid only for the one-soliton case.
The integral is taken over the space-like plane $t=const$ and does not
depend on $t$.

Using (\ref{ccreat}) and (\ref{ccreat2}) and replacing 
$\int\!\diff{x}\diff{y}\,\tr
\to 2\pi\theta\,\mbox{Tr} \equiv 2\pi\theta\,\mbox{Tr}_{\Hcal_t}$,
we can calculate the energy (\ref{energy}) of the abelian soliton
$\Phi=1+\frac{\mu-{\bar\mu}}{{\bar\mu}}P_0$:
\begin{align}
E(\mu)|_{\scriptscriptstyle U(1)}\ &=\ \frac14
\frac{(\mu{-}{\bar\mu})({\bar\mu}{-}\mu)(1{+}\mu{\bar\mu})^2}{(\mu{\bar\mu})^2}
\cdot \frac{4}{i(\mu{-}\bar{\mu}{-}\mu^{-1}{+}\bar{\mu}^{-1})}\cdot
2\pi\bet\,\mbox{Tr}\,(\pa_w P_0\,\pa_{\wb}P_0+\pa_{\wb}P_0\,\pa_w P_0)
\nonumber\\[4pt]
&=\ \frac{2\pi(\bar{\mu}-\mu)(1+\mu{\bar \mu})}{i\,\mu{\bar \mu}}\cdot
\underbrace{\mbox{Tr}\,[c\,,\,P_0]\,[P_0\,,\,\cdag]\,}_{\scriptstyle =1}
\ =\ 8 \pi\gam \sin^2 \varphi \quad .
\end{align}
To arrive at the last expression, we have performed the polar decomposition
\begin{equation}
\mu\ =\ \alf\,e^{-i\varphi} \qquad\textrm{and defined}\qquad
\gam\ =\ \frac{1+\alf^2}{2\alf\,\sin\varphi} \quad.
\end{equation}
The velocity of the soliton in the $xy$ plane is (cf. \cite{ward})
\begin{equation}\label{velo}
(v_x,v_y)\ =\
-\Bigl(\frac{2\alf\cos\varphi}{\alf^2+1}\ ,\ \frac{\alf^2-1}{\alf^2+1}\Bigr)
\quad,\qquad\textrm{so that}\qquad\gam^{-1}\ =\ \sqrt{1-v_x^2-v_y^2}\quad.
\end{equation}

We repeat the computation for the $U(2)$ soliton built with~(\ref{mat}).
To this end we differentiate
$\Phi=1+\frac{\mu-{\bar\mu}}{{\bar\mu}}P$ with the help
of (\ref{ccreat2}) and make repeated use of the identities
$\ f(N)\,c=c\,f(N{-}1)\ $ and
$\ \cdag f(N)=f(N{-}1)\,\cdag$.
Substituting into (\ref{energy}), we obtain
\begin{align}
E(\mu)|_{\scriptscriptstyle U(2)}\ 
&=\ \frac{2\pi(\bar{\mu}-\mu)(1+\mu{\bar \mu})}{i\,\mu{\bar \mu}}\cdot
\mbox{Tr}\,[c\,,\,P]\,[P\,,\,\cdag] \nonumber \\[8pt]
&=\ 8\pi \gam \sin^2 \varphi\ \mbox{Tr}\, 
\Biggl| \left(\begin{matrix}
c\,\frac{-\l\lb}{(N-1+\l\lb)(N+\l\lb)} & 
\frac{\l\,\l\lb}{(N+\l\lb)(N+1+\l\lb)} \\[8pt]
c\,c\,\frac{-\lb}{(N-1+\l\lb)(N+\l\lb)} &
c\,\frac{\l\lb}{(N+\l\lb)(N+1+\l\lb)}
\end{matrix}\right) \Biggr|^2
\nonumber \\[6pt]
&=\ 8\pi \gam \sin^2 \varphi \ \mbox{Tr}\, \biggl\{
\frac{(\l\lb)^2}{(N{+}\l\lb)(N{+}1{+}\l\lb)^2} + 
\frac{\l\lb\;N}{(N{-}1{+}\l\lb)(N{+}\l\lb)^2} \biggr\} 
\nonumber \\[6pt]
&=\ 8\pi \gam \sin^2 \varphi \ \sum_{n=0}^{\infty} 
\frac{\l\lb}{(n{+}\l\lb)(n{+}1{+}\l\lb)} \ =\
8 \pi \gam \sin^2 \varphi \quad,
\end{align}
where some care must be taken for $|\l|=1$.
We see that the energies of the one-soliton solutions 
for the groups $U(1)$ and $U(2)$ are finite 
and depend on the velocity (\ref{velo}) of the solitons.  
Notice that the energies do not depend on~$\theta$ and
are identical to the commutative result for~$SU(2)$~\cite{ward}.

In order to obtain a solution describing $m$ solitons moving with the 
{\it same\/} velocity one should use in (\ref{toned}) two polynomials in~$c$ 
of degree~$m$. More general multi-soliton solutions featuring {\it relative\/}
motion of the individual lumps can be constructed by using
$m$ {\it different\/} matrices $T_k$ and applying formulae 
(\ref{movesol2})-(\ref{explsol2}) with~(\ref{gensol}). 
All these {\it moving\/} solitons have a brane-world interpretation: 
Since $m$-soliton solutions of an effective $U(n)$ Yang-Mills-Higgs theory
in $2{+}1$ dimensions correspond to $m$ $D0$-branes inside $n$ coincident 
$D2$-branes, we see that our solitons describe $D0$-branes {\it moving\/} 
inside the $D2$-branes.

\section{Solitons in 1+1 and instantons in 2+0 dimensions}
\noindent
Consider the effective field theory on the $D1$-brane world volume briefly 
discussed in section~\ref{effective}. After changing field variables via 
$\ \pa_x \phi= \Phi^{-1}\pa_t\Phi\ $ and $\ \pa_t\phi=\Phi^{-1}\pa_x\Phi$ 
its equation of motion takes the form 
%\vspace{-0.5cm}
\begin{equation}\label{61}
\pa_x(\Phi^{-1}\pa_x\Phi)\ -\ \pa_t(\Phi^{-1}\pa_t\Phi)\ =\ 0 \quad .
\end{equation}
This is the field equation of the standard sigma model in $1{+}1$ dimensions
with the field $\Phi$ taking values in the group $U(n)$.

Obviously our choice (\ref{bchoice}) of the $B$-field leads to commutative 
$\R^{1,1}$, where $f{\star}g=f\,g$, 
and there appears no star product in (\ref{61}) (nor in (\ref{d1a})).
In principle one can turn on a more general $B$-field background with nonzero 
`{\it electric\/}' components $B_{14}=-B_{41}$. 
In~\cite{susskind} this type of field has been considered in 
much detail, and it was shown that it does not admit a zero-slope limit 
which produces a field theory on a noncommutative space-time. Thus, 
open \N2 strings only yield the sigma model on {\it commutative\/} $\R^{1,1}$.
In order to derive solitonic solutions of its equations (\ref{d1a}) 
and (\ref{61}) one should simply drop the $y$~dependence from all formulae 
of section~\ref{noncomm} (i.e. impose $\pa_y=0$ on all fields). 
No additional work is necessary here.

Let us finally 
have a look at the effective field theory on the {\it Euclidean\/} 
$D1$-brane world volume. It is easy to see that after the field redefinition 
$\ \pa_x\phi=\Phi^{-1}\pa_y\Phi\ $ and $\ \pa_y\phi= -\Phi^{-1}\pa_x\Phi\ $ 
its equation of motion becomes
%\vspace{-0.5cm}
\begin{equation}\label{62}
\pa_x(\Phi^{-1}\star\pa_x\Phi)\ +\ \pa_y(\Phi^{-1}\star\pa_y\Phi)\ =\ 0\quad.
\end{equation}  
This equation describes the $U(n)$ sigma model on the {\it noncommutative\/} 
Euclidean space $\R^{2,0}$. It is interesting to note that these equations can 
be derived by reduction of the noncommutative self-dual Yang-Mills equations 
on~$\R^{2,2}$ but not from their Euclidean variant on~$\R^{4,0}$ 
(cf.~\cite{IP}).
Solutions of (\ref{62}) are easily obtained from the solutions in $2{+}1$ 
dimensions formulated in section~\ref{noncomm}. One should simply put 
$\mu_k=-i\ \forall k$ in all formulae. 
Then $w_k=x{+}iy=z$ and ${\bar w}_k=x{-}iy=\zb$, i.e. instanton solutions of 
noncommutative two-dimensional sigma models arise as 
{\it static\/} solutions of the modified sigma model in $2{+}1$ dimensions.
If in (\ref{onedimsol}) we consider projectors of rank one, we get the $\C P^n$
chiral model on a noncommutative plane. Solutions of its field equation were
studied in~\cite{lee}.

\medskip
\noindent
{\large{\bf Acknowledgements}}

\smallskip
\noindent
The work of O.L. and A.D.P. is partially supported 
by DFG under grant Le~838/7-1.
B.S. is grateful to the `Studienstiftung des deutschen Volkes' 
for support.

\end{document}